\documentclass{article}

\usepackage{PRIMEarxiv}

\usepackage[utf8]{inputenc} 
\usepackage[T1]{fontenc}    
\usepackage{hyperref}       
\usepackage{url}            
\usepackage{booktabs}       
\usepackage{amsfonts}       
\usepackage{nicefrac}       
\usepackage{microtype}      
\usepackage{lipsum}
\usepackage{fancyhdr}       
\usepackage{graphicx}       
\graphicspath{{media/}}     

\title{Rethinking Technological Readiness in the Era of AI Uncertainty
}

\author{
  S. Tucker Browne \\
  United States Air Force \\
  Washington, DC, USA \\
   \And
  Mark M. Bailey \\
  National Intelligence University \\
  Washington, DC, USA\\
}

\begin{document}
\maketitle

\begin{abstract}
Artificial intelligence (AI) is poised to revolutionize military combat systems, but ensuring these AI-enabled capabilities are truly mission-ready presents new challenges. We argue that current technology readiness assessments fail to capture critical AI-specific factors, leading to potential risks in deployment. We propose a new AI Readiness Framework to evaluate the maturity and trustworthiness of AI components in military systems. The central thesis is that a tailored framework – analogous to traditional Technology Readiness Levels (TRL) but expanded for AI – can better gauge an AI system’s reliability, safety, and suitability for combat use. \cite{NASA2023} Using current data evaluation tools and testing practices, we demonstrate the framework’s feasibility for near-term implementation. This structured approach provides military decision-makers with clearer insight into whether an AI-enabled system has met the necessary standards of performance, transparency, and human integration to be deployed with confidence, thus advancing the field of defense technology management and risk assessment.
\end{abstract}


\section{Introduction}

Advances in artificial intelligence (AI) promise enhanced capabilities for military combat systems, from autonomous drones to decision-support algorithms.\cite{Scharre2018} These benefits come with new risks: AI systems can behave unpredictably, lack transparency, and perform inconsistently outside of controlled settings.\cite{Bailey2025} To overcome these challenges, a dedicated AI Readiness Framework is needed to systematically assess whether AI-enabled military systems are truly prepared for deployment. This article contends that defense organizations should adopt an AI-specific readiness assessment, analogous to (but more comprehensive than) traditional metrics like Technology Readiness Levels (TRLs),\cite{NASA2023} to ensure justified confidence in AI systems before they are fielded. We begin by examining the limitations of current readiness assessment metrics (such as TRLs) when applied to AI. We then introduce a new framework with specific criteria designed to evaluate AI system maturity, explaining our rationale for each criterion and discussing implementation considerations.\cite{Bailey2024} Next, we analyze how the proposed framework addresses critical AI system challenges, including “hallucinations,” lack of explainability, and performance variability in operational scenarios. Finally, we outline the framework’s applicability to current military AI programs and conclude with recommendations for integrating this approach into defense technology management.

\section{Background: AI in Military Combat Systems and the Readiness Challenge}

\textit{Defining AI Systems in Scope:} In this discussion, an AI-enabled military system refers to narrowly focused AI and machine learning applications embedded in combat platforms and decision processes. These include, for example, computer vision for target recognition, autonomous navigation for unmanned vehicles, sensor fusion and situational awareness tools, decision-support systems that suggest tactics or logistical actions, and natural language processing systems for intelligence analysis. Such AI systems are data-driven (often employing neural networks or other machine learning models) and perform specific tasks under human supervision, rather than possessing broad general intelligence. Notably, these AI modalities correspond to core military functions: perception (e.g., image classifiers spotting targets), planning/control (e.g., an autonomous route-finding algorithm for a robotic vehicle), prediction (e.g., maintenance failure predictors), and decision aids (e.g., recommender systems for commanders). This analysis focuses on evaluating narrow AI capabilities integral to combat systems. We do not address speculative artificial general intelligence; instead, the framework targets AI subsystems likely to be fielded in the near-term, each with distinct readiness considerations.

\textit{The Limits of Traditional Readiness Metrics:} The Department of Defense (DoD) and other agencies have long used Technology Readiness Levels\cite{NASA2023} (TRLs) to measure the maturity of new technologies. TRLs provide a scale from 1 (basic principles observed) to 9 (actual system proven in operational environment) to indicate how developed and tested a technology is. However, TRLs were designed primarily for hardware systems and emphasize technical integration milestones. When applied to AI, traditional TRL assessments fall short in several ways. Notably, AI performance can degrade unpredictably in novel scenarios – a phenomenon not captured by simple lab prototype demonstrations. An AI model might achieve TRL 7 (prototype in operational environment) yet still fail when facing inputs outside its training distribution. Moreover, TRLs do not explicitly account for data quality, model transparency, or human trust, which are critical for AI. As a result, programs might declare an AI component “ready” based on integration testing, while overlooking unresolved algorithmic risks. Indeed, a recent national security commission observed that legacy test and evaluation methods are “not sufficient” for AI systems and that agencies “lack common metrics to assess trustworthiness that AI systems will perform as intended.” In short, current readiness paradigms provide an incomplete picture of an AI system’s true deployment readiness.

\section{Unique Challenges of AI-Enabled Systems}
Several characteristics of AI systems make readiness assessment more complex than for traditional military technologies. Key challenges include:

\begin{itemize}
\item \textit{Unpredictable Behavior and “Hallucinations”:} Advanced AI models, especially large neural networks and generative AI, can produce outputs that are unexpected or incorrect without warning. For instance, large language models may generate plausible-sounding but false information – a problem known as hallucination.\cite{Rawte2023} In mission contexts, an AI that occasionally outputs incorrect target identifications or faulty recommendations can have dire consequences.\cite{HSTodayAIControl} This unpredictability complicates confidence in readiness: decision-makers worry that if an AI is routinely unreliable, then leaders will not adopt such systems.  A readiness framework must therefore evaluate whether an AI system’s performance is consistent and whether any tendency to hallucinate or error is sufficiently mitigated.

\item \textit{Lack of Explainability (Transparency):} Many AI algorithms, particularly deep learning models, act as “black boxes,” meaning their internal decision-making logic is not transparent to users or even developers. Military leaders and operators need to understand why an AI system produces a recommendation or action, especially in high-stakes scenarios.\cite{Transparency2023} Complex neural networks may contain millions of parameters interwoven in ways that defy human comprehension; tracing how an input leads to an output can be functionally impossible. Even if full explainability isn’t achievable (as many models are inherently complex), the framework should consider the level of “black box” risk an AI brings.\cite{Sullivan2024} Indeed, experts note that lack of transparency and explainability is one of the greatest concerns with advanced AI capabilities. Therefore, an AI Readiness assessment must consider whether the system provides adequate transparency or rationale for its outputs.

\item \textit{Inconsistent Performance Across Scenarios (Robustness):} AI systems can be highly brittle – performing extremely well on data similar to their training set but faltering in edge cases or new environments. A target-recognition AI might excel in clear weather but fail in fog or against adversarial camouflage. Unlike mechanical systems, which typically fail in predictable and repeatable ways, AI failures can be context-dependent and hard to anticipate. Studies have shown that current AI models “are often brittle when operating at the edges of their performance competence, and it is difficult to anticipate their competence boundaries”.\cite{NSCAI} This brittleness means that an AI which passes tests in one scenario might not generalize to others. Traditional TRL exercises (e.g., a single field demonstration) might not reveal these weak points. Therefore, assessing AI Readiness requires much more extensive scenario-based testing and evaluation of robustness – how well the AI maintains performance when conditions change or when facing inputs crafted to confuse it (e.g., adversarial inputs or unexpected data). A framework must account for whether such stress testing has been done and how the AI handles variability.

\item \textit{Data Dependence and Quality Issues:} The performance and reliability of AI systems are fundamentally tied to the quality of data used to train and test them.\cite{Ntoutsi2020} If the training data are biased, incomplete, or not representative of real operating conditions, the AI’s behavior will reflect those gaps. An AI might perform well in development but encounter entirely new data patterns in a real conflict (for example, a vision AI trained mostly on desert terrain may struggle when deployed in jungle terrain). Data problems can also manifest as vulnerabilities, e.g., the system might be easily fooled by poisoned or false data. Traditional readiness metrics do not examine the suitability of the training data or the adequacy of testing data coverage. Thus, an AI readiness evaluation should include a data readiness component, ensuring that the system’s training and validation datasets are sufficient in size, diversity, and accuracy for the intended mission. It should also consider whether mechanisms are in place to update or retrain the model as new data becomes available or as adversaries adapt.

\item \textit{Human-System Integration and Trust:} Finally, the “readiness” of an AI capability is not only about the technology itself but also about the human operators, commanders, and processes around it.\cite{Hancock2023} A highly advanced AI targeting assistant, for example, is useless if operators are not trained to use it effectively or if they distrust its recommendations and thus ignore them. Past experiences with automation in military systems have shown that humans can both over-trust (placing blind faith in automation) and under-trust (rejecting its advice outright) depending on how the system is presented and what feedback it provides. Thus, part of readiness is human readiness: Do the end-users understand the AI’s capabilities and limitations? Have procedures been developed to appropriately incorporate the AI’s input into decision-making? Is there a clear doctrine on human oversight of the AI (for instance, when it’s allowed to act autonomously versus when direct human intervention is required)? The DoD’s adoption of Responsible AI principles and human-machine teaming research emphasizes the importance of AI that complements human decision-making and maintains the “continued centrality of human judgment”.  Therefore, an AI Readiness Framework should explicitly evaluate the status of training, doctrines, and user trust for that system.

\end{itemize}

In summary, AI-enabled military systems bring challenges of unpredictable errors, opaque reasoning, context-specific failures, data risks, and complex human integration requirements. The problem we address is how to assess, in a structured way, whether these challenges have been sufficiently mitigated for a given AI system intended for deployment. The next section introduces a proposed AI Readiness Framework intended to fill this gap, by incorporating criteria that speak to these AI-specific issues while building on the concept of staged readiness similar to TRLs.

\section{Proposed AI Readiness Framework}

To close the assessment gap, we propose an AI Readiness Framework as a structured approach to evaluate AI-enabled combat systems. This framework extends the familiar notion of readiness levels by adding qualitative criteria addressing AI’s unique dimensions of risk and performance. The framework is inspired by TRLs, in that it envisions staged milestones from initial development to field deployment, but it augments those milestones with five key readiness criteria focused on AI concerns. The proposed criteria are: (1) Alignment, (2) Justified Confidence, (3) Governance, (4) Data Readiness, and (5) Human Readiness. Each criterion represents a critical facet of an AI system’s maturity and suitability for use. An AI system would be assessed against all five criteria to determine if it meets the minimum acceptable standard in each, akin to “gates” it must pass through on the way to full operational deployment. By evaluating these dimensions, commanders and acquisition professionals can identify areas needing improvement and ensure no aspect of AI risk is overlooked before fielding the system.

Below, we define and justify each criterion, discuss its rationale and importance, and note any limitations or trade-offs it introduces. We also consider how each can be measured with current or anticipated data and tools, assessing the framework’s feasibility for practical adoption.

\subsection{Alignment}

\textit{Definition:} Alignment refers to the degree to which the AI system’s objectives, behavior, and outputs are consistent with the intended goals, ethical constraints, and rules of engagement set by human commanders.\cite{Ji2023} In simple terms, an aligned AI will do what we want it to do – and nothing counter to our intent – even in complex or novel situations.

\textit{Rationale and Importance:} Alignment is arguably the foremost criterion because a misaligned AI can become not just a neutral tool failure but an active liability. In military operations, alignment means the AI’s actions or recommendations must remain within the bounds of lawful orders, mission objectives, and ethical considerations (such as avoiding civilian harm). Beyond traditional performance metrics, it involves ensuring the AI will not pursue an unintended goal or interpret its task in a problematic way. For example, an AI controlling a loitering munition must recognize and respect no-strike zones or cease-fire conditions – it should not “creatively” seek targets outside its authorization. Alignment also encompasses the mitigation of bias in AI decision-making. If an intelligence analysis AI is aligned with commanders’ priorities, it should not consistently skew or omit information due to biases in its training data. Ensuring alignment addresses the concern that AI systems could otherwise yield outcomes that are tactically effective but strategically or ethically unacceptable. It also ties directly to the DoD’s principle that AI use must remain accountable to human judgment and legal standards.\cite{DoDRAI}

\textit{Addressing AI Challenges:} This criterion helps tackle the risk of AI hallucinations or rogue behavior by imposing checks that the AI’s outputs adhere to validated truth or authorized bounds.\cite{BreakingDefense2024} An aligned AI language model, for instance, would be constrained (via training or rule-based filters) not to fabricate information outside of known data, thereby reducing hallucinations. Alignment criteria would also incorporate validation tests for ethical compliance – essentially red-teaming the AI with various scenarios to see if it ever produces forbidden actions or recommendations. If it does, it fails alignment readiness. While alignment testing is not guaranteed to catch every possible failure (AI behavior in truly unprecedented situations might defy prediction), it sets a standard that obvious misalignments are discovered in advance. In practice, emerging techniques like adversarial scenario generation and reward modeling for alignment are tools that can be applied here, making this criterion increasingly feasible to evaluate with current AI safety research advances.\cite{Amodei2016}

\textit{Limitations and Trade-offs:} Measuring alignment is challenging. It is partly a normative judgment – what counts as “aligned” enough? Complete assurance of alignment is arguably impossible, especially for an AI deployed in an open-ended environment. Thus, the framework would likely define alignment thresholds (for example, the AI passes alignment readiness if extensive testing shows no violation of certain constraints). There is also a trade-off: heavily constraining an AI to ensure alignment may reduce its effectiveness or require simplified model designs. For instance, adding rigorous rule-based checks to a learning algorithm might prevent creative solutions – this could be a pro (safer behavior) but also a con (potentially less optimal performance). Nevertheless, given that unaligned AI behavior could be catastrophic, the bias in the framework is towards caution.

\textit{Feasibility:} With current tools, alignment can be partially assessed using structured scenario tests, ethical checklists, and compliance with known directives. As AI governance tools mature, we anticipate more automated alignment auditing (for example, software that scans AI decision logs for signs of objective function deviation). In short, alignment is a qualitative but critical readiness gate: an AI system should not be deployed until commanders are satisfied it will pursue the right objectives in the right way.

\subsection{Justified Confidence}

\textit{Definition:} Justified Confidence refers to the evidence-based assurance that the AI system will perform its intended function reliably and within acceptable risk margins. This criterion is about trust grounded in testing and validation – having rigorous proof that the system works as designed under expected conditions (and understanding its failure modes). It encompasses classical performance metrics (accuracy, error rates) but goes further to include confidence in the AI’s behavior under stress, uncertainty, and in interaction with other systems or humans.

\textit{Rationale and Importance:} The term “justified confidence” is drawn from defense AI literature emphasizing that operators and leaders need well-founded trust in AI, not blind faith. For an AI system to be ready, it is not enough that developers are confident – that confidence must be justified by transparent data from evaluations. This criterion essentially asks: Has the AI been tested enough, and do the results justify trusting it in battle? It stresses robust Test and Evaluation, Verification and Validation of the AI. Traditional TRLs might consider whether a prototype demonstration was successful, but justified confidence demands statistical and empirical evidence of performance across a range of scenarios. This includes measuring false positive/negative rates, failure frequency, and the impact of uncertainties. The National Security Commission on Artificial Intelligence (NSCAI) Final Report underscores that achieving acceptable AI performance often involves understanding and accepting certain levels of risk.\cite{NSCAI} Our framework makes that explicit by requiring decision-makers to evaluate whether the demonstrated performance (and its bounds) are acceptable for the mission at hand. If an AI has a 5\% error rate in ideal conditions and perhaps higher in edge cases, is that risk tolerable in context (e.g., is it controlling a lethal weapon or just providing recommendations)? Only if the answer is “yes, with good reason” do we have justified confidence.

\textit{Addressing AI Challenges:} This criterion directly addresses inconsistent performance and brittleness. To have justified confidence, one must test the AI in varied conditions, including edge cases. Suppose we have an autonomous vehicle AI. Justified confidence suggests it has been test-driven through as many different environments and adversarial conditions as feasible, and it handles them to the required standard. The framework would encourage practices like aggressive stress testing and red-teaming of AI (as recommended by the NSCAI to uncover where the AI might break. It also inherently covers reliability over time – continuous monitoring and validation results should be gathered to ensure the model doesn’t drift from its original performance). An important aspect is that justified confidence includes performance with humans in the loop. For example, if an AI works well in isolation but confuses human operators in a team setting, we cannot be confident in its real-world performance. Thus, this criterion overlaps with human readiness by also requiring that any human-AI team dynamics don’t degrade performance. In essence, justified confidence is the accumulation of all testing evidence to conclude the AI will perform as intended. By insisting on this, the framework mitigates the risk of deploying AI systems that pass basic checks but fail under pressure.

\textit{Limitations and Trade-offs:} One limitation is that obtaining “complete” confidence is impossible – there will always be unknown unknowns. The framework doesn’t demand perfection but justification: documented testing, clear understanding of the AI’s capabilities and limits. Another challenge is the resource and time intensity of extensive AI testing. Unlike a physical system that might require few test conditions, an AI may need thousands of simulated sorties or test cases to build statistical confidence, especially for safety-critical functions. This can slow down deployment and incur costs (e.g., a trade-off of caution vs. speed). However, given the potential high cost of AI failure in combat, this trade-off is warranted. Modern tools like simulation environments, digital twins, and automated test harnesses make broad testing more feasible than before, and techniques for formal verification of certain AI behaviors are improving.

\textit{Feasibility:} Today, the DoD is investing in specialized test infrastructure for AI (for instance, the Joint AI Test Infrastructure \cite{CDAOAI} mentioned by CDAO). These provide the kind of data pipelines and validation suites that can underpin justified confidence assessments. So, while challenging, this criterion can be met with current technology by dedicating sufficient T\&E effort and using advanced analytics to measure AI reliability.

\subsection{Governance}

\textit{Definition:} Governance in the AI Readiness context refers to the organizational and process measures in place to oversee the AI system’s development, deployment, and lifecycle management. This includes compliance with ethical guidelines, legal requirements, and safety standards; the presence of accountability structures; version control and configuration management of the AI models; and plans for continuous monitoring and improvement once the AI is fielded. Essentially, governance ensures that the AI system is embedded in a framework of rules and oversight that manages its risks over time.

\textit{Rationale and Importance:} Governance is a criterion often absent in technical readiness discussions, but it is crucial for AI because of the dynamic and potentially evolving nature of these systems. A machine learning model might change if retrained on new data; its performance might drift. Without strong governance, an AI that was ready at deployment can become “un-ready” later. Moreover, military use of AI is guided by policies (such as the DoD’s Responsible AI Principles\cite{DoDRAI} and international law commitments) that must be adhered to throughout the system’s life. Governance readiness means the program has done things like: established an AI risk management plan, conducted independent audits or peer reviews of the AI, ensured there is a fallback or “kill switch” if the AI misbehaves, and set up channels for operators to report issues. It also examines whether the procurement and contracting included necessary clauses about AI performance and data rights (so that the military isn’t stuck with an opaque model it can’t update or inspect). In short, governance ensures that deploying the AI system won’t outpace the institution’s ability to control it. This fosters long-term trust: people are more confident in AI that comes with visible oversight and accountability. For example, if a unit knows there is a clear protocol for what to do if the AI gives a questionable output (and that someone is monitoring AI decisions at higher echelons), they might be more willing to use it.

\textit{Addressing AI Challenges:} The governance criterion addresses explainability and safety issues indirectly by enforcing process discipline and accountability. For instance, a governance-ready AI program would have documentation and traceability for its training data and algorithms, aligning with the idea of making AI “traceable” or at least documented for outside scrutiny. This can help mitigate the black-box issue, not by magically explaining the AI’s internals, but by ensuring the developers provide as much transparency as possible (e.g., model cards, known limitations, intended use conditions). Governance also covers how the AI’s outputs are used: are there standard operating procedures requiring human review of AI-generated targets or recommendations? If so, even if the AI has a minor hallucination, governance processes can catch it before it does harm (for example, an intelligence AI might be required to have an analyst vet its report before it influences decisions). Additionally, proper governance means the system is fielded with monitoring hooks – logging performance, detecting anomalies, and triggering re-evaluation if needed. This continuous oversight is critical because it acknowledges the fact that we cannot predict everything in testing. Instead, we set up a safety net during operations. The framework’s governance check would confirm such a safety net exists.

\textit{Limitations and Trade-offs:} Focusing on governance might be seen as bureaucratic by some. It can introduce additional paperwork and slower approvals (e.g., waiting for an ethics review or external audit before deployment). However, governance measures are deliberate frictions to ensure caution with powerful AI capabilities. A limitation is that governance quality can be hard to quantify – it’s certainly possible to have formal processes that are merely box-checking. The framework would need to define concrete governance indicators (like “an independent test organization has certified the AI against X standards” or “an AI safety board reviewed the system and issued recommendations, which were implemented”). Another challenge is keeping governance agile: overly rigid controls might stifle rapid iteration on AI systems, which is problematic given the fast pace of AI improvements. The framework thus calls for governance but not stagnation – ideally a balance where oversight exists but can react quickly (for instance, expedited processes in wartime that still include essential checks).

\textit{Feasibility:} Implementing governance readiness is largely about policy and organizational action, which the DoD is already pursuing. The DoD CDAO’s Responsible AI initiative \cite{CDAOAI}, for example, is creating toolkits to ensure “justified confidence in AI-enabled systems” through governance mechanisms. We anticipate that by following emerging standards (like NIST’s AI Risk Management Framework or NATO AI governance guidelines \cite{NATOAI}), programs can meet this criterion. It might involve additional training for program managers and commanders on AI risk governance, which is a current gap but one that is being acknowledged and addressed across defense communities.

\subsection{Data Readiness Level (DRL)}

\textit{Definition:} Data Readiness evaluates whether the data underpinning the AI system is sufficient and appropriate for the task. This includes the quality, quantity, diversity, and relevance of the training data, as well as the availability of ongoing data pipelines for maintenance. In our framework, a high Data Readiness Level (DRL) means the AI has been trained on a comprehensive, representative dataset and tested on operationally relevant test cases, with processes in place to obtain new data as conditions evolve.

\textit{Rationale and Importance:} Data is the fuel of AI. In military AI projects, one of the most common failure points is not the algorithm itself but the data used to train it. If the data are poor, even the most advanced AI will under-perform or behave unpredictably. A readiness assessment that ignores data would miss fundamental issues – for instance, hidden biases that could cause the AI to misidentify targets of a certain type because those were underrepresented in training. By explicitly rating data readiness, we prompt developers to treat data as a first-class citizen in system development. This means asking questions like: Have we gathered enough real-world data or high-fidelity simulated data? Does it cover the range of scenarios the AI will face (day/night, different terrains, different adversary tactics)? How have we verified and cleaned the data? Are there known gaps? The DRL concept also extends to test data: an AI’s performance should be evaluated on test scenarios that reflect real mission conditions, not just on the training set or trivial cases. Moreover, the framework would look at whether the program has the infrastructure to continually feed the AI new data from the field (for updates or retraining) – crucial for long-lived systems where the operational environment might shift. Emphasizing data readiness aligns the framework with the reality that AI development is often 80\% data preparation and 20\% coding; a system is only as ready as its data allows.

\textit{Addressing AI Challenges:} By assessing DRL, the framework encourages collection of diverse data sets to avoid brittleness. For example, if a drone’s vision AI is only trained on summer imagery, its DRL would be low for winter operations; acknowledging that fact forces remedial action (acquire winter data). Data readiness also addresses some aspects of hallucination for generative models: if a language model is intended to provide factual briefings, a high DRL would imply it was trained on verified military domain knowledge and perhaps connected to a database – reducing its propensity to hallucinate untruths because it has a solid factual grounding. Regarding explainability, while data quality alone doesn’t make an AI more explainable, having well-curated data can eliminate spurious correlations that make models act weirdly. It also contributes to alignment: if you want the AI aligned with human values, the training data should include human-supervised examples of correct behavior. In essence, DRL is a preventive metric: ensuring the AI has “seen” what it needs to see before deployment, thus preempting many failure modes.

\textit{Limitations and Trade-offs:} One limitation is that obtaining highly representative data for every conceivable scenario can be extremely difficult, especially for future combat environments that haven’t been encountered yet. There’s a risk of diminishing returns – collecting ever more data with only marginal gains. Also, some military data are scarce or expensive to produce (e.g., live-fire combat data for obvious reasons). The framework, however, doesn’t mandate perfection, but awareness: if data are lacking, that should be explicitly noted and mitigation strategies should be in place. A trade-off here is that insisting on data readiness might delay projects while data are gathered or force reliance on synthetic data. Synthetic or simulated data can help but may not capture reality perfectly, which itself is a risk if over-relied upon. Thus, the framework might give a moderate DRL to a system heavily trained on simulation, with a requirement to validate on real data as soon as possible.

\textit{Feasibility:} Measuring DRL can be done now by evaluating the training dataset statistics and test results. There are proposed quantitative metrics (like coverage of input feature space, data quality indexes, etc.) that programs can adopt. Organizations are also increasingly aware of data needs – for instance, the creation of large military image datasets or war-game scenario libraries to train AI. So, incorporating a DRL check is practical, and it compels programs to either acquire needed data or clearly state the limits of their data (which then informs commanders of what situations might be risky for the AI).

\subsection{Human Readiness Level (HRL)}

\textit{Definition:} Human Readiness assesses the preparedness of the human elements – the end users, operators, commanders, and support personnel – to effectively work with the AI system. This includes training and education about the AI, the development of tactics, techniques, and procedures (TTPs) for its use, user interface design and usability, and the degree of trust/calibration users have with the AI’s outputs. We can analogize this to a Human Readiness Level (HRL), complementing the technical readiness of the AI with the readiness of the humans in the loop or on the loop.

\textit{Rationale and Importance:} No military capability exists in a vacuum; even autonomous systems operate within human command structures. An AI might meet all technical criteria but still fail in the field because operators misuse it or decision-makers misunderstand its reports. Therefore, human readiness is essential. This criterion forces evaluators to ask: Have the users been trained to interpret the AI’s recommendations? Do they know its limits and failure modes? Is there doctrine on how to deploy and supervise the AI? For example, if a command staff receives a recommendation from an AI planning tool, do they treat it as one input among many (as intended), defer to it blindly, or ignore it entirely? Each of those outcomes is possible, and only training and proper integration can lead to the desired middle ground where AI is used appropriately. Additionally, human readiness looks at things like interface design – is the AI’s output presented in a clear way? Does the operator get explanation or confidence levels from the AI to inform their decisions? If the AI raises an alert, does the operator know what action to take? This criterion is very much about human-systems integration, a domain of research that has taken on new urgency with AI advances. The concept of human-AI teaming is relevant: a system is ready not when the AI alone is good, but when the human-AI team can accomplish the mission effectively. Thus, the HRL check is about ensuring the people and procedures are as ready as the technology.

\textit{Addressing AI Challenges:} Human readiness is crucial for mitigating the explainability and trust issues. Rather than expecting a perfect explanation from the AI, we ensure the human understands as much as needed to trust the system within bounds. Training can help users grasp the AI’s “mental model” (even if the algorithm is complex, analogies and experience can teach operators when to rely on it vs. when to be skeptical). For instance, pilots working with an AI co-pilot (like an autonomous wingman drone) need training sorties with the system to build appropriate trust. Studies have found that without deliberate effort, humans might either over-trust an AI (leading to complacency and failure to catch its mistakes) or under-trust it (leading to refusal to use a potentially life-saving capability). By measuring HRL, we require evidence that such training and calibration have occurred: e.g., field exercises where the operators demonstrated correct use of the AI, or surveys indicating operators understand the AI’s accuracy rates and limitations. Human readiness also includes ensuring leadership buy-in – if commanders don’t trust the AI, they may under-utilize it. Conversely, over-hyping AI could lead to misuse. A balanced approach is part of readiness. In summary, this criterion ensures that issues of explainability and trust are managed not just by the AI’s design, but by user education and organizational culture. It also ensures that a plan exists for human oversight: if the AI system does something unusual, are humans ready and empowered to intervene or shut it down? If not, the system shouldn’t be considered fully ready.

\textit{Limitations and Trade-offs:} Assessing human readiness can be subjective. It may rely on qualitative judgments (e.g., an evaluator might interview service members about their comfort with the AI). However, there are ways to formalize it: training completion rates, performance in exercises, or even simulation-based certification where operators must respond correctly in AI-assisted scenarios. Another limitation is timing: human readiness often can only be fully achieved when the AI is near deployment, as it requires the actual system (or a high-fidelity simulator) to train on. This means HRL might lag behind other criteria. A trade-off here is that pushing an AI out quickly without extensive user training could risk incidents, whereas waiting for all users to be fully trained might delay fielding a useful capability. The framework would recommend at least a minimal HRL (for example, key personnel trained and initial TTPs defined) before first deployment, and then continuing to raise human readiness in parallel with early operational use.

\textit{Feasibility:} The military has established training pipelines and simulation environments that can be leveraged. For instance, if an AI is added to a fighter aircraft, part of testing should involve pilots training with it in simulators – metrics from those sessions (like mission success rates or pilot feedback scores) can gauge HRL. Doctrine and manuals can be drafted as part of the acquisition process to guide appropriate use. None of this is outside current capability; it simply needs to be recognized as a formal part of readiness. In fact, treating human readiness as equal to technical readiness is common in fields like aviation safety – we insist pilots be certified on new systems. The same must hold for AI: the “crew” of an AI-enabled system (which might be an operator or could be the developers maintaining the model) must be certified as ready. The framework thus makes human readiness an explicit checkpoint.

\textit{Summary of Framework Criteria:} An AI-enabled system can be said to reach full readiness (analogous to TRL 9) when it scores satisfactorily on all five dimensions: its goals are aligned with command intent, there is justified confidence from rigorous testing, governance and oversight mechanisms are in place, the data foundation is solid, and the human operators are prepared and integrated. If any one of these is lacking, the framework would signal that deployment carries unaddressed risk in that area. This multi-criteria approach ensures a holistic assessment. Importantly, the framework does not promise to eliminate all AI risks; it brings them to light so that informed decisions can be made. In some cases, a system might deploy even if one criterion is only partially met (due to urgent need), but the shortfall would be known and mitigation strategies adopted. In other cases, a low score on one criterion (say alignment or confidence) should be a show-stopper until corrected, for safety reasons. This structured checklist is itself a benefit: it forces program managers to document how they have dealt with issues like explainability or data bias – topics that might otherwise be overlooked.

\section{Addressing Critical AI Challenges with the Framework}

A primary motivation for this new framework is to better account for AI-specific challenges. Here we explicitly consider how the proposed criteria engage with the issues of hallucination, lack of explainability, and inconsistent real-world performance, as often raised by skeptical commanders. We also acknowledge where the framework has limits in fully resolving these problems.

\textit{Mitigating Hallucinations and Unpredictable Outputs:} In our framework, the Alignment and Justified Confidence criteria work together to account for hallucinations and nonsensical outputs. Alignment criteria ensure the AI is designed and constrained to stay truthful and on-mission. For instance, an aligned AI chatbot for military use would be connected to verified databases and have rules against making up facts, an approach known to reduce hallucinations. Meanwhile, justified confidence requires thorough testing of the AI’s outputs. If an AI consistently shows a 10\% hallucination rate in trials (similar to what researchers observed with some models) the framework would flag it as not ready. Only with improvements – such as integrating a verification module or simplifying the model – to bring that error rate down to an acceptable level would the system clear the confidence criterion. In practice, this might involve using techniques like Retrieval Augmented Generation (which was mentioned as reducing hallucinations to 0.1\% in tests). The framework thus does not eliminate the possibility of hallucination, but it refuses to green-light a system until that possibility is minimized and measured. It also ensures stakeholders are aware of any remaining hallucination risk. For example, if an AI assistant for intelligence analysis still might produce an odd incorrect statement occasionally, commanders would be informed (via the readiness assessment) that its Alignment/Confidence is imperfect and that human analysts must double-check certain outputs. Therefore, hallucinations are addressed to the extent current technology allows – primarily by requiring alignment techniques and extensive validation. What the framework cannot promise is absolute elimination of unpredictable outputs; instead, it ensures they are either driven to a low probability or the system is not deployed.

\textit{Improving Explainability and Transparency:} The framework addresses explainability and transparency through the Governance and Human Readiness criteria. Governance pushes for documentation, standards, and possibly the use of "explainable AI (XAI)" methods.\cite{DARPA_XAI} A system that uses simpler models or includes explanation modules (like saliency maps for image AI to highlight what it “saw”) would score better in readiness than an inscrutable black box with equal performance. Moreover, governance would ensure that even if the model itself isn’t fully interpretable, the deployment includes measures like requiring a human analyst to interpret and contextualize the AI’s output for commanders. On the human side, the framework demands training so operators understand the AI’s logic as much as possible. For instance, if the AI is a neural network for target identification, operators might be trained on the kinds of features the network uses and its known failure cases, effectively giving them a mental model of the AI. This goes a long way to bridging the explainability gap – the AI might not be transparent internally, but its behavior patterns are known to the user. The framework also explicitly acknowledges if an AI has low explainability: during the readiness assessment, this would appear as a risk (perhaps under Governance: “Model is black-box; no \textit{post-hoc} explainers implemented”). If that risk is not mitigated (by additional tools or human procedures), the system might not pass readiness. In cases where a high-performing but non-explainable AI is deemed necessary, the framework at least forces a conscious decision: leadership must waive the explainability requirement knowing the downsides. In summary, the framework addresses explainability as far as possible by requiring either technical or procedural solutions for transparency. It doesn’t magically make opaque AI clear, but it ensures the issue is not ignored. We also admit that not all AI decisions will be fully explainable – this is a known open problem – but the framework’s stance is that unexplained intelligence is inherently untrustworthy. Thus, an AI with serious explainability issues may be marked as not fully ready, or only ready for limited use, which incentivizes developers to incorporate explainability techniques from the start.

\textit{Ensuring Robust Performance Across Real-World Scenarios} AI systems often struggle when reality diverges from training conditions. Our Justified Confidence and Data Readiness criteria directly respond to this by emphasizing diverse testing and data coverage. To illustrate, consider inconsistent performance: a vision AI that functions in lab lighting but not work at dusk. Under our framework, this shortcoming would be revealed either by reviewing the training data (Data Readiness would note the model was trained mostly on daytime images, lowering its DRL for night operations) or by testing (Justified Confidence trials would show poor results at dusk, failing the performance assurance). The remedy would be to gather more dusk images and retrain (data improvement) and then pass additional tests. In effect, the framework’s process is iterative: if an AI fails to generalize in a certain scenario, it is not ready – but this failure tells developers exactly what to fix to improve readiness (get new data or adjust the model). Additionally, the Alignment criterion can indirectly support consistency: a well-aligned AI might be designed to recognize when it’s out of its depth and defer to a human or default safe mode, rather than press on with a guess. This kind of behavior can prevent catastrophic failures in unanticipated situations and would be credited in a readiness evaluation. It’s worth noting that no framework can guarantee performance in truly unforeseen conditions (war is the realm of surprise, after all). However, by requiring rigorous stress testing and scenario diversity, we greatly reduce the “unknown unknowns.” We also incorporate the idea of continuous monitoring (from Governance) – meaning if the AI encounters a new scenario in the field, its performance is logged and reviewed. If a new failure mode appears, that triggers a reassessment. In sum, the framework makes the AI’s consistency (or lack thereof) visible and actionable. Systems that demonstrate consistent performance across a wide range get the green light; those that are inconsistent are either held back or deployed with strict caveats (like “do not use in X condition”), which again ensures commanders are aware of limitations.

By explicitly focusing on these challenges, the proposed framework offers a more nuanced readiness evaluation. It does not fully resolve all AI issues – some, like deep explainability or absolute assurance of no hallucination, remain research problems. However, it integrates current best practices to manage these challenges: enforce alignment to minimize off-target behavior, demand evidence for reliability, integrate humans to catch what AI misses, and govern the system to respond to issues. In areas where the framework cannot solve a challenge, it demands transparency about it. For example, if after all efforts an AI still has a 1\% unexplainable error rate, the framework would not hide this; it would be documented in the readiness report so that risk is accounted for in operational planning. This is a significant improvement over legacy approaches that might simply certify a system as ready without ever addressing these AI-centric concerns.

\section{Implementation Considerations and Feasibility}

Translating the AI Readiness Framework from concept to practice will require effort, but it is feasible with current and emerging tools. Each criterion can leverage existing methodologies:

\textit{Alignment:} Implementing alignment checks could use red-teaming exercises, where experts attempt to find scenarios that cause the AI to misbehave. Techniques from AI safety research, such as adversarial testing and reinforcement learning with human feedback, can be applied during development to steer the AI toward aligned behavior. The framework’s alignment assessment can be informed by the results of such exercises and by compliance checklists (e.g., ensuring the AI meets the DoD’s ethical AI guidelines).\cite{DoDRAI} We foresee that as the defense community builds experience, standardized alignment evaluation protocols will emerge – similar to how cybersecurity has penetration testing. In the interim, alignment can be qualitatively assessed by panels of subject matter experts reviewing the AI’s design and behavior logs for signs of goal misalignment.

\textit{Justified Confidence (Testing \& Evaluation):} There is a growing suite of tools for AI test and evaluation. Simulation environments (for drones, vehicles, etc.) allow repeated and safe testing of AI in virtual combat scenarios. Moreover, the DoD’s AI test infrastructure programs aim to provide common test datasets and metrics. For example, an AI targeting algorithm might be tested against a standardized set of images or scenarios and must achieve a certain accuracy with narrow uncertainty bounds. Formal verification techniques, while limited, can be used for specific components (especially for simpler AI or rule-based parts of the system) to mathematically prove certain properties (like “will not fire unless target is classified as hostile with >99\% confidence”). In terms of process, programs will need to allocate more time in their schedule for extensive AI testing, potentially including field trials under contested conditions. This is a shift, but not an unreasonable one – analogous shifts happened when software became a big part of systems (introducing software testing phases). The key is to integrate AI testing early and continuously; tools for automated testing (like continuous integration pipelines that re-run test suites whenever the model is updated) can help manage this without excessive burden.

\textit{Governance:} Ensuring governance readiness is largely about adopting and enforcing policies. This might involve creating an “AI Readiness Review Board” in acquisition programs, much like flight readiness reviews for aircraft. Such a board would verify that documentation is complete (e.g., data origin documented, model version control in place), that an appropriate legal review of the AI’s use has been done (especially if autonomy in weapons is involved, complying with directives like DODD 3000.09 \cite{DODD3000}), and that there are responsible individuals assigned for ongoing monitoring. Technologically, governance can be facilitated by tracking systems – for example, AI model management platforms that log every change made to a model and its performance impact. Feasibly, the DoD could mandate that any AI component in a system must come with an “AI safety case” report before deployment, akin to a safety certification. The NIST AI Risk Management Framework \cite{NISTAI} provides templates for identifying and controlling risks which align with governance checks; utilizing such frameworks can operationalize the governance criterion. The main challenge here is cultural and procedural, not a lack of tools.

\textit{Data Readiness:} There are already concepts of data readiness levels in some AI projects. Tools for data profiling and validation exist – for instance, one can compute statistics on dataset diversity, or use AI itself to scan for bias in data labels. Synthetic data generation tools (like simulation engines or generative models) can fill identified gaps. The framework would prompt programs to explicitly plan for data: e.g., if an AI needs to be trained on rare events, the program might invest in a high-fidelity simulator or leverage historical data from allies if available. Feasibility is generally good here, as long as data are treated as part of the procurement (for example, ensuring contracts specify delivery of training data or rights to collect operational data). The biggest hurdle would be if critical data are classified or hard to obtain; in such cases, the readiness assessment might remain low until that data can be incorporated, perhaps limiting initial deployment to environments similar to what was tested. Over time, the accumulation of operational data will improve DRL, and the framework encourages that feedback loop.

\textit{Human Readiness:} Implementing this will require coordination with training commands and doctrine developers. Feasible steps include adding modules about the new AI system in the training curricula of operators, creating simulation scenarios that include the AI for units to practice, and writing or updating field manuals and tactics guides to include the AI’s role. Surveys and user studies can measure trust and understanding; these can be executed by human factors teams that the military often employs for new equipment (for example, when a new aircraft or software system is introduced, human factors assessments are routine). The main requirement is to not treat the AI system as just a piece of equipment, but as a capability that changes how people work – thus requiring training and possibly even organizational changes (like maybe having an “AI officer” role to monitor AI outputs in a headquarters). The framework’s HRL can be made concrete by milestones such as: X\% of operators qualified on the system, a tabletop exercise conducted with leadership to gauge decision integration, or a trial deployment where the AI is used in a limited capacity and feedback collected. All these are feasible with current practice; it just extends the concept of test trials and training to the AI domain.

\textit{Data and Tool Support:} Fortunately, many of the framework’s assessments can be supported by analytical tools. For example, dashboards could be created for each AI system in development, showing progress on each readiness criterion: number of test cases passed (Confidence), percent of required data collected (Data Readiness), training hours completed (Human Readiness), etc. This quantification helps manage the process and gives decision-makers a snapshot of readiness. Moreover, an important aspect of feasibility is who performs the assessment. Ideally, an independent evaluation team (perhaps within the test community or CDAO office) would apply the AI Readiness Framework to each project, much like independent test evaluators do for hardware. This third-party assessment ensures objectivity and helps overcome optimism bias from developers. The independent team can utilize all the tools and methods discussed to produce an AI Readiness report. There may be initial growing pains as both programs and evaluators learn how to measure things like alignment or human trust, but iterative improvement is expected.

In summary, implementing the AI Readiness Framework is doable with today’s technology and organizational structures, especially if there is top-level support for making AI assurance a priority. The framework largely repurposes existing best practices (from software engineering, testing, training, etc.) under a unified umbrella. It is also scalable: it could be applied in lightweight form for low-risk AI (e.g., a logistics AI might not need as exhaustive a process) or in full rigor for high-risk AI (e.g., autonomous weapons). By starting to apply these criteria now on pilot programs, the defense community can refine the process and build a knowledge base of what works best, paving the way for a formal adoption of such frameworks in acquisition policy.

\section{Conclusion}

The integration of AI into military combat systems demands a rethinking of how we evaluate readiness and risk. This article presented a structured AI Readiness Framework as a solution to the shortcomings of traditional readiness assessments when applied to artificial intelligence. We assert that without such a framework, we risk deploying AI capabilities that are technically operable but not truly prepared for the complex, high-stakes environments of warfare. By introducing criteria for Alignment, Justified Confidence, Governance, Data Readiness, and Human Readiness, we expand the evaluation beyond mere technical functionality to include ethical, organizational, and reliability considerations. We justified each criterion and showed how, collectively, they address critical AI challenges: reducing the chance of AI systems acting unpredictably, highlighting the need for explainability and user trust, and ensuring robust performance across varied scenarios.

This framework, aligned with the spirit of Technology Readiness Levels \cite{NASA2023} but adapted for AI’s unique demands, provides military decision-makers a more nuanced “checklist” before declaring an AI system mission-ready. It preserves most of the content of existing assessments – such as requiring demonstration in operational conditions – while adding depth in areas like data quality and human factors. In doing so, it does not fully eliminate all the uncertainties of AI; instead, it brings those uncertainties to light and imposes a higher standard of evidence and oversight. A system that might have been waved through under a conventional process must, under this framework, earn trust on multiple fronts. We have also outlined how current data, tools, and processes can feasibly implement the framework, indicating that this is not a distant ideal but an achievable near-term improvement.

Ultimately, the AI Readiness Framework is about responsibility in innovation: harnessing cutting-edge AI for defense while rigorously managing the risks it entails. As AI technologies continue to evolve, this framework can be updated – new criteria can be added if needed (for example, if future AI systems have reinforcement learning capabilities in the field, we may add a criterion for on-line learning safety). But the core insight remains that AI Readiness is multi-dimensional. The hope is that by institutionalizing such an approach, the military will avoid tragic surprises and instead field AI that servicemembers and commanders can trust with their lives and in their missions. In an era where AI capabilities are rapidly advancing, ensuring readiness in the fullest sense will be a key differentiator between those who integrate AI successfully and those who fall victim to its pitfalls. The framework proposed here is a step toward the successful, safe, and effective adoption of artificial intelligence in national security.

\vspace{1em}

\small{\textit{Disclaimer: The authors are responsible for the content of this article. The views expressed do not reflect the official policy or position of the National Intelligence University, the Department of Defense, the Office of the Director of National Intelligence, the U.S. Intelligence Community, or the U.S. Government.}

\small{\textbf{Maj. S. Tucker Browne} is highly involved in the acquisition and test world for the U.S. Air Force. He is a developmental fighter test pilot with over 1400 flight hours, a distinguished graduate of the \href{https://www.edwards.af.mil/Units/USAFTPS/}{USAF Test Pilot School}, and a graduate of the \href{https://ni-u.edu/}{National Intelligence University}, where his thesis focused on the assessment of military artificial intelligence systems. In his current assignment, he works at the Pentagon as a program manager for multiple joint developmental programs.}

\small{\textbf{Dr. Mark Bailey} writes about the intersection between artificial intelligence, complexity, and national security. His work has been featured in notable publications such as the journal Futures, Nautilus, The Conversation, and Homeland Security Today, and he was named to \href{https://trailblazers.hstoday.us/mark-bailey/}{Homeland’s 50 Trailblazers of 2023}. Mark is the Department Chair for Cyber Intelligence and Data Science, as well as the Director of the Data Science Intelligence Center, at the \href{https://ni-u.edu/}{National Intelligence University}, and is a Senior Fellow for AI, Complexity and Risk at the \href{https://www.fau.edu/future-mind/}{Center for the Future Mind}. He is the author of \href{https://www.unknowableminds.com}{\textit{Unknowable Minds: Philosophical Insights on AI and Autonomous Weapons}}. Mark is also a Lieutenant Colonel the U.S. Army Reserve.}}



\bibliographystyle{unsrt}  
\bibliography{references}

\begin{thebibliography}{10}

\bibitem{NASA2023}
Catherine~G. Manning.
\newblock Technology readiness levels.
\newblock Space Communications and Navigation Program, NASA, 9 2023.

\bibitem{Scharre2018}
Paul Scharre.
\newblock {\em Army of None: Autonomous Weapons and the Future of War}.
\newblock W. W. Norton, New York, 2018.

\bibitem{Bailey2025}
Mark Bailey.
\newblock {\em Unknowable Minds: Philosophical Insights on AI and Autonomous Weapons}.
\newblock Imprint Academic, Exeter, UK, 2025.

\bibitem{Bailey2024}
Samuel Browne, Thomas Pike, and Mark~M. Bailey.
\newblock A proposed framework for artificial intelligence safety and technology readiness assessments for national security applications.
\newblock OSF Preprints, 8 2024.

\bibitem{Rawte2023}
Vipula Rawte, Amit Sheth, and Amitava Das.
\newblock A survey of hallucination in large foundation models.
\newblock {\em arXiv preprint}, 2023.

\bibitem{HSTodayAIControl}
{Homeland Security Today}.
\newblock Artificial intelligence, critical systems, and the control problem, 2022.
\newblock Accessed: April 11, 2025.

\bibitem{Transparency2023}
Nagadivya Balasubramaniam, Marjo Kauppinen, Antti Rannisto, Kari Hiekkanen, and Sari Kujala.
\newblock Transparency and explainability of ai systems: From ethical guidelines to requirements.
\newblock {\em Information and Software Technology}, 159, 2023.

\bibitem{Sullivan2024}
Scott Sullivan.
\newblock Targeting the black box: The need to reprioritize ai explainability.
\newblock Articles of War (Lieber Institute, West Point), 8 2024.

\bibitem{NSCAI}
{National Security Commission on Artificial Intelligence (NSCAI)}.
\newblock Chapter 7: Establishing justified confidence in ai systems, 3 2021.

\bibitem{Ntoutsi2020}
Eirini Ntoutsi, Pavlos Fafalios, Ujwal Gadiraju, Vasileios Iosifidis, Wolfgang Nejdl, Maria-Esther Vidal, Salvatore Ruggieri, Franco Turini, Symeon Papadopoulos, Emmanouil Krasanakis, Ioannis Kompatsiaris, Katharina Kinder-Kurlanda, Claudia Wagner, Fariba Karimi, Miriam Fernandez, Harith Alani, Bettina Berendt, Tina Kruegel, Christian Heinze, Klaus Broelemann, Gjergji Kasneci, Thanassis Tiropanis, and Steffen Staab.
\newblock Bias in data-driven artificial intelligence systems—an introductory survey.
\newblock {\em Wiley Interdisciplinary Reviews: Data Mining and Knowledge Discovery}, 10(3):e1356, 2020.

\bibitem{Hancock2023}
Peter~A. Hancock, Joseph~L. Lyons, Deborah~A. Boehm-Davis, Raja Parasuraman, and John~D. Lee.
\newblock Trust, workload, and performance in human–artificial intelligence teaming: A meta-analysis.
\newblock {\em Journal of Mechanical Design}, 147(1):011702, 2023.

\bibitem{Ji2023}
et~al. Jiaming~Ji.
\newblock Ai alignment: A comprehensive survey.
\newblock {\em arXiv preprint}, 2023.

\bibitem{DoDRAI}
{United States Department of Defense}.
\newblock Dod adopts ethical principles for artificial intelligence, 2 2020.

\bibitem{BreakingDefense2024}
{Breaking Defense Staff}.
\newblock For artificial intelligence to be mission critical, it must hallucinate less.
\newblock Breaking Defense, 12 2024.

\bibitem{Amodei2016}
Dario Amodei, Chris Olah, Jacob Steinhardt, Paul Christiano, John Schulman, and Dan Mané.
\newblock Concrete problems in ai safety.
\newblock {\em arXiv preprint}, 2016.

\bibitem{CDAOAI}
{Chief Digital and Artificial Intelligence Office (CDAO)}.
\newblock Testing and evaluation, 2025.
\newblock Accessed March 25, 2025.

\bibitem{NATOAI}
{North Atlantic Treaty Organization (NATO)}.
\newblock Summary of nato's revised artificial intelligence (ai) strategy, 7 2024.

\bibitem{DARPA_XAI}
{Defense Advanced Research Projects Agency (DARPA)}.
\newblock Explainable artificial intelligence (xai), 2023.
\newblock Accessed: April 11, 2025.

\bibitem{DODD3000}
{U.S. Department of Defense}.
\newblock Directive 3000.09: Autonomy in weapon systems, 2017.
\newblock Updated May 8, 2017.

\bibitem{NISTAI}
{National Institute of Standards and Technology (NIST)}.
\newblock Ai test, evaluation, validation, and verification (tevv), 2025.
\newblock Accessed March 25, 2025.

\end{thebibliography}

\end{document}